\documentclass[aps,prl,twocolumn,superscriptaddress]{revtex4}
\newcommand{\PRE}[1]{}       % Use if journal style
\usepackage{bm}
\usepackage{epsfig}

\newcommand{\postscript}[2]{\setlength{\epsfxsize}{#2\hsize}
   \centerline{\epsfbox{#1}}}

\newcommand{\mstar}{M_{\ast}}
\newcommand{\mbh}{M_{\text{BH}}}
\newcommand{\mbhmin}{M_{\text{BH}}^{\text{min}}}

\newcommand{\gev}{\text{GeV}}
\newcommand{\tev}{\text{TeV}}
\newcommand{\pb}{\text{pb}}

\newcommand{\km}{\text{km}}
\newcommand{\s}{\text{s}}

\newcommand{\cmwe}{\text{cmwe}}
\newcommand{\kmwe}{\text{kmwe}}
\newcommand{\xmax}{X_{\text{max}}}

\newcommand{\eg}{{\em e.g.}}

\newcommand{\eqref}[1]{Eq.~(\ref{#1})}

\begin{document}

\preprint{
\hfil
\begin{minipage}[t]{3in}
\begin{flushright}
\vspace*{.4in}
MIT--CTP--3182\\
UCI--TR--2001--27\\
UK/01--07\\
hep-ph/0109106
\end{flushright}
\end{minipage}
}

\title{
\PRE{\vspace*{1.5in}}
Black Hole Production by Cosmic Rays}

\author{Jonathan L.~Feng}
\affiliation{Center for Theoretical Physics, Massachusetts Institute
of Technology, Cambridge, MA 02139
\PRE{\vspace*{.1in}}
}
\affiliation{Department of Physics and Astronomy, University of
California, Irvine, CA 92697
\PRE{\vspace*{.1in}}
}

\author{Alfred D.~Shapere%
\PRE{\vspace*{0.3in}}
}
\affiliation{Center for Theoretical Physics, Massachusetts Institute
of Technology, Cambridge, MA 02139
\PRE{\vspace*{.1in}}
}
\affiliation{Department of Physics, University of Kentucky,
Lexington, KY 40502
\PRE{\vspace*{.5in}}
}

%\date{September 2001}

\begin{abstract} 
Ultra-high energy cosmic rays create black holes in scenarios with
extra dimensions and TeV-scale gravity.  In particular, cosmic
neutrinos will produce black holes deep in the atmosphere, initiating
quasi-horizontal showers far above the standard model rate.  At the
Auger Observatory, hundreds of black hole events may be observed,
providing evidence for extra dimensions and the first opportunity for
experimental study of microscopic black holes. If no black holes are
found, the fundamental Planck scale must be above 2 TeV for any number
of extra dimensions.

\vspace*{.05in}
\centerline{MIT--CTP--3182, UCI--TR--2001--27, UK/01--07}
\end{abstract}

%\pacs{04.70.Dy, 96.40.Tv, 13.15.+g, 04.50.+h}
%04.70.Dy   Quantum physics of black holes
%96.40.Tv   Neutrino and muon cosmic rays
%13.15.+g   Neutrino interactions
%04.50.+h   Gravity in more than four dimensions

\maketitle

Black holes are among the most captivating and inaccessible phenomena
in physics.  In principle, tiny black holes can be produced in
particle collisions with center-of-mass energies above the Planck
scale $\mstar$, where they should be well described semiclassically
and thermodynamically~\cite{Hawking:1975sw}.  However, in conventional
four-dimensional theories, $\mstar \sim 10^{19}~\gev$.  Given
currently accessible energies $\alt 1~\tev$, the study of such black
holes is far beyond the realm of experimental particle physics.

In models with extra dimensions, however, the fundamental Planck scale
may be much lower. If this is the case, black hole production and
evaporation might be observed in particle
collisions~\cite{Banks:1999gd,Emparan:2000rs}.  Beginning later this
decade, for example, the Large Hadron Collider (LHC) at CERN will
begin operation with parton center-of-mass energies of several TeV.
Assuming $M_*$ of order 1 TeV, the authors of
Refs.~\cite{Giddings:2001bu,Dimopoulos:2001hw} have noted the
distinctive characteristics of black hole production and find event
rates as large as $10^8$ per year.

As high as it is, the energy range probed by the LHC is modest
compared to that of ultra-high energy cosmic rays, which have been
observed to interact in the Earth's atmosphere with center-of-mass
energies in excess of 100 TeV.  As we will see, cosmic neutrinos with
energies above $10^6~\gev$ are particularly effective sources of black
holes, with production cross sections as much as two or more orders of
magnitude above standard model (SM) predictions. These black holes
decay rapidly, initiating spectacular quasi-horizontal showers deep in
the atmosphere.

Observation of such showers at the rates we predict would be a strong
indication of new TeV-scale physics.  In the SM, while Earth-skimming
ultra-high energy neutrinos may be observed with reasonable rates by
fluorescence detectors~\cite{Feng:2001ue} and ground
arrays~\cite{Bertou:2001vm}, those that pass only through the
atmosphere are extremely difficult to detect.  Even at the large
Pierre Auger Observatory, SM interactions are expected to produce only
a fraction of an event per
year~\cite{Capelle:1998zz,Coutu:1999ub,Diaz:2001mv}.  In contrast,
with conservative neutrino flux estimates, we find that Auger could
detect hundreds of black holes before the LHC begins operation,
providing evidence for TeV-scale gravity and extra dimensions and
making possible the experimental study of black holes in the late
stages of Hawking evaporation.

Low-scale gravity may be realized if the conventional four spacetime
dimensions are supplemented by $n$ additional spatial dimensions. SM
matter and gauge fields are typically assumed to be confined to the
four dimensions of our world.  However, gravity propagates in the full
(4+$n$)-dimensional space with Einstein action
\begin{equation}
S_E = \frac{1}{8 \pi} \mstar^{2+n} \int d^{4+n}x \sqrt{-g} \,
\hbox{$1\over 2$} {\cal R} \ ,
\end{equation}
where $\mstar$ is the fundamental Planck scale.  If the
$n$-dimensional space is flat and compact with volume
$V_n$~\cite{Antoniadis:1998ig}, the observed gravitational strength is
reproduced provided $\mstar^{2+n} V_n \approx (1.2\times
10^{19}~\gev)^2$.  For large $V_n$, $\mstar$ near the TeV scale is
possible: current bounds are $\mstar \agt 40~\tev$ for $n=2$, $\mstar
\agt 400~\gev$ for $n=4$, and $\mstar \agt 300~\gev$ for $n =
6$~\cite{Peskin:2000ti}.  For $n=1$, $\mstar \sim \tev$ is excluded in
this context.  However, in alternative scenarios with warped
metrics~\cite{Randall:1999ee}, gravity may become strong below
$1~\tev$, even for $n=1$~\cite{Davoudiasl:2000jd}.  Our discussion
will not depend on the details of these scenarios, as long as the
black holes produced are smaller than the compactification radii
(curvature scales) in flat (warped) scenarios, and so well
approximated by (4+$n$)-dimensional asymptotically Minkowskian
solutions.

In theories with TeV-scale gravity, many effects may alter cosmic ray
interactions with observable consequences~\cite{Nussinov:1999jt}.
However, in contrast to all other processes, our understanding of
black hole properties is expected to be qualitatively sound for
center-of-mass energies above $\mstar$ and increasingly valid the
farther above $\mstar$ one probes.  Black holes, then, may provide a
uniquely reliable probe of extra-dimensional effects above $\mstar$,
where the full range of cosmic ray energies may be exploited.

The Schwarzschild radius for a (4+$n$)-dimensional, neutral,
non-spinning black hole with mass $\mbh$ is~\cite{Myers:1986un}
\begin{equation}
r_s (\mbh^2) = \frac{1}{\sqrt{\pi} \mstar} 
\Biggl[ \frac{\mbh}{\mstar} \Biggr]^{\frac{1}{1+n}} 
\left[ \frac{8 \Gamma \left(\frac{3+n}{2} \right)}
{2+n} \right]^{\frac{1}{1+n}} \ .
\end{equation}
Black hole formation is expected when partons $i$ and $j$ with
center-of-mass energy $\sqrt{\hat{s}}$ pass within a distance
$r_s(\hat{s})$, suggesting a geometrical cross section of order
\begin{equation}
\label{sigmabh}
\hat{\sigma} (ij \to \text{BH}) (\hat{s}) \approx \pi r_s^2(\hat{s}) \ .
\end{equation}
We will take this as an adequate approximation and assume that a black
hole of mass $\mbh = \sqrt{\hat{s}}$ is formed.  (Numerical analysis
of classical head-on collisions in four dimensions finds $\mbh \approx
0.8 \sqrt{\hat{s}}$~\cite{D'Eath:1993gr}.)  The suppression factor of
Ref.~\cite{Voloshin:2001vs} has been
disputed~\cite{Dimopoulos:2001qe}; we have not included it here.  The
neutrino-nucleon scattering cross section is then
\begin{equation}
\sigma ( \nu N \to \text{BH}) = \sum_i \int_{(\mbhmin{})^2/s}^1 dx\,
\hat{\sigma}_i ( xs ) \, f_i (x, Q) \ ,
\end{equation}
where $s = 2 m_N E_{\nu}$, the sum is over all partons in the nucleon,
the $f_i$ are parton distribution functions (pdfs), and $\mbhmin$ is
the minimal black hole mass for which \eqref{sigmabh} is expected to
be valid.  We set momentum transfer $Q = \min \{ \mbh, 10~\tev \}$,
where the upper limit is from the CTEQ5M1 pdfs~\cite{Lai:2000wy};
$\sigma ( \nu N \to \text{BH})$ is insensitive to the details of this
choice.  For the conservative fluxes considered below, our results are
also rather insensitive to $x < 10^{-5}$.  For concreteness, however,
we extrapolate to $x < 10^{-5}$ assuming $f_i(x,Q) \propto
x^{-[1+\lambda_i(Q)]}$.  Finally, we choose $\mbhmin = \mstar$.  The
relatively mild dependence on $\mbhmin$ is discussed below.

\begin{figure}[tb]
\postscript{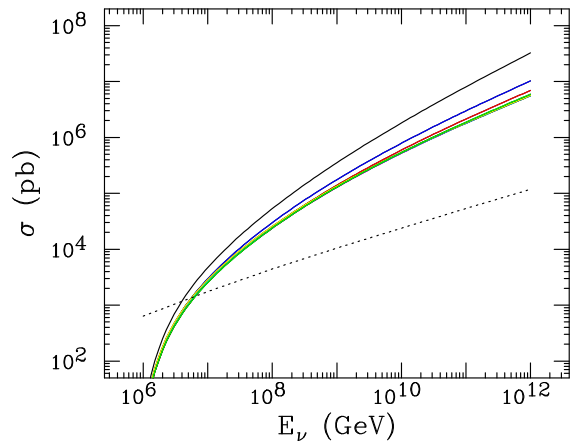}{0.86}
%\postscript{sig1.eps}{0.66}
\caption{Cross sections $\sigma ( \nu N \to \text{BH})$ for $\mstar =
\mbhmin = 1~\tev$ and $n=1, \ldots, 7$ from above. (The last four
curves are virtually indistinguishable.)  The dotted curve is for the
SM process $\nu N \to \ell X$.}
\label{fig:sig1}
\end{figure}

Cross sections for black hole production by cosmic neutrinos are given
in Fig.~\ref{fig:sig1}. The SM cross section for $\nu N \to \ell X$ is
included for comparison.  In contrast to the SM process, black hole
production is not suppressed by perturbative couplings and is enhanced
by the sum over all partons, particularly the gluon.  In addition,
while the SM cross section grows rapidly with $E_{\nu}$, as is well
known, the black hole cross section grows even more rapidly: for large
$n$, it has the asymptotic behavior $\sigma \propto
E_{\nu}^{\lambda_i(10~\tev)} \approx E_{\nu}^{0.45}$.  As a result of
these effects, black hole production may exceed deep inelastic
scattering rates by two or more orders of magnitude.

Although greatly reduced by black hole production, neutrino
interaction lengths $L=1.7\times 10^7~\kmwe~(\pb/\sigma)$ are still
far larger than the Earth's atmospheric depth, which is only
$0.36~\kmwe$ even when traversed horizontally.  Neutrinos therefore
produce black holes uniformly at all atmospheric depths.  As a result,
the most promising signal of black hole creation by cosmic rays is
quasi-horizontal showers initiated by neutrinos deep in the
atmosphere.  At these angles, the likelihood of interaction is
maximized and the background from hadronic cosmic rays is eliminated,
since these shower high in the atmosphere.  The number of black holes
detected is, then,
\begin{equation}
{\cal N} = \int dE_{\nu}\, N_A \, \frac{d\Phi}{dE_{\nu}} \,  
\sigma(E_{\nu}) \, A(E_{\nu}) \, T \ ,
\label{numevents}
\end{equation}
where $A(E_{\nu})$ is a given observatory's acceptance for
quasi-horizontal showers in cm$^3$ water equivalent steradians
(cm$^3$we sr), $N_A = 6.022 \times 10^{23}$ is Avogadro's number,
$d\Phi/dE_{\nu}$ is the source flux of neutrinos, and $T$ is the
running time of the detector.

\begin{figure}[tbp]
\postscript{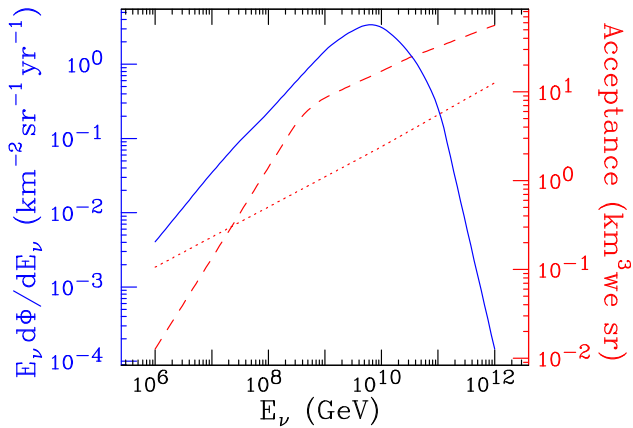}{0.96}
%\postscript{fluxaccept.eps}{0.76}
\caption{Neutrino flux from Greisen photoproduction
(solid)~\protect\cite{Stecker:1979ah}, and ground array
(dashed)~\protect\cite{Capelle:1998zz} and fluorescence
(dotted)~\protect\cite{Diaz:2001mv} acceptances of one Auger site for
quasi-horizontal hadronic showers.  For fluorescence detection, a duty
cycle of 10\% has been included. }
\label{fig:fluxaccept}
\end{figure}

There are many possible sources of ultra-high energy neutrinos.  Here
we conservatively consider only the `guaranteed' flux of Greisen
neutrinos produced by interactions of the observed ultra-high energy
cosmic rays with the cosmic microwave
background~\cite{Greisen:1966jv}.  This flux is subject to
uncertainties; we adopt the results of Ref.~\cite{Stecker:1979ah},
shown in Fig.~\ref{fig:fluxaccept}.  The flux estimates of
Refs.~\cite{Hill:1985mk} produce similar event rates, while the strong
source evolution case of Ref.~\cite{Yoshida:1997ie} enhances the
results below by over an order of magnitude.  New physics might also
increase the neutrino flux.  In particular, many proposed explanations
of cosmic rays with energies above the Greisen-Zatsepin-Kuz'min
cutoff~\cite{Greisen:1966jv,Zatsepin:1966jv} would boost these event
rates by several orders of magnitude.

Quasi-horizontal showers may be observed by air shower ground arrays
or air fluorescence detectors.  The largest near-future cosmic ray
experiment is the Auger Observatory, a hybrid detector consisting of
two sites, each with surface area $3000~\km^2$.  Construction of the
southern site is in progress, with a counterpart planned in the
northern hemisphere.  Auger acceptances for deeply penetrating air
showers have been studied in
Refs.~\cite{Capelle:1998zz,Coutu:1999ub,Yoshida:1997ie,Diaz:2001mv}.
Black holes decay thermally, according to the number of degrees of
freedom available, and so their decays are mainly
hadronic~\cite{Giddings:2001bu,Dimopoulos:2001hw}.  We therefore
consider the hadronic shower acceptance for ground arrays, including
`partially contained' showers~\cite{Capelle:1998zz}.  For
fluorescence, we use the results of Ref.~\cite{Diaz:2001mv} for
showers with zenith angles above $60^{\circ}$ initiated at depths
greater than $1250~\cmwe$.  These acceptances are given in
Fig.~\ref{fig:fluxaccept}. A duty cycle of 10\% has been included for
fluorescence, where observations are limited to cloudless, moonless
nights. At $E_{\nu} \sim 10^{10}~\gev$ where the Greisen flux peaks,
the ground array is more sensitive, and so we focus on ground array
rates below.  Note, however, that future detectors, such as Telescope
Array and the space-based OWL and EUSO, will improve this fluorescence
acceptance by one to three orders of magnitude~\cite{Stecker:2000ek}.

\begin{figure}[tbp]
\postscript{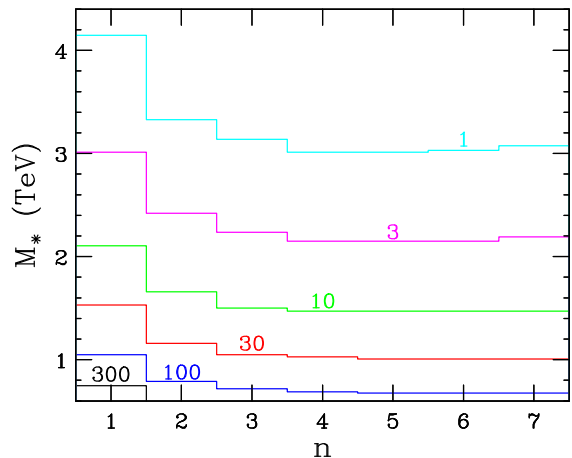}{0.86}
%\postscript{n.eps}{0.66}
\caption{The number of black holes detected by the ground array in 5
Auger site-years as a function of $\mstar=\mbhmin$ and the number of
extra dimensions $n$.}
\label{fig:n}
\end{figure}

Given the cross sections of Fig.~\ref{fig:sig1} and the flux and
acceptances of Fig.~\ref{fig:fluxaccept}, the number of events is
determined by \eqref{numevents}.  The results for ground arrays are
given in Fig.~\ref{fig:n}.  Tens to hundreds of events are possible
for $\mstar \approx 1~\tev$.  Tens of black holes may also be detected
by fluorescence.  For larger $\mstar$, $\hat{\sigma} (ij \to
\text{BH})$ falls rapidly as $\mstar^{-(4+2n)/(1+n)}$.  Nevertheless,
requiring 3 events for discovery, black hole production probes Planck
scales as high as 3 TeV for $n=1$, and 2 TeV for all $n$.  If no
events are seen, barring a neutrino flux significantly below our
conservative estimate, a stringent lower bound of $\mstar \agt 2~\tev$
may be set for all $n$.

\begin{figure}[tbp]
\postscript{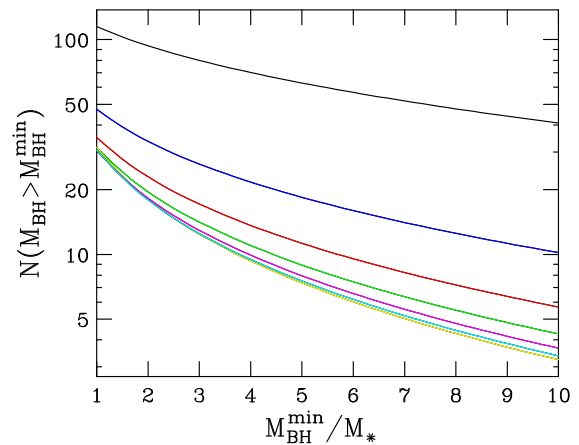}{0.86}
%\postscript{nmbhmin.eps}{0.66}
\caption{The number of black holes detected by the ground array in 5
Auger site-years as a function of $\mbhmin$ for $\mstar=1~\tev$ and
$n=1, \ldots, 7$ from above. }
\label{fig:nmbhmin}
\end{figure}

The results of Fig.~\ref{fig:n} are for $\mbhmin = \mstar$.  While the
semiclassical approximation is invalid for $\mbh \approx \mstar$, this
is a calculational, not physical, limitation and does not imply that
production of black holes or similar states in this mass range is
suppressed~\cite{Dimopoulos:2001qe}; in fact, it may just as well be
enhanced.  Nevertheless, it is comforting to know that our results are
not strongly sensitive to this assumption.  In Fig.~\ref{fig:nmbhmin},
the dependence on $\mbhmin$ is shown.  For $\mbhmin = 5 \mstar$, event
rates are reduced by factors of 2 for $n=1$ and 4 for large $n$.
While these reductions are substantial, they are extremely mild
relative to the case at colliders.  At the LHC, the requirement $\mbh
> 5 \mstar$ suppresses event rates by factors of a hundred or
more~\cite{Dimopoulos:2001hw}.  For cosmic rays, while the black hole
mass distribution is still peaked at low masses as a result of
enhancements from pdfs at low $x$, the reduction is far more modest.

If an anomalously large quasi-horizontal shower rate is found, it may
be identified as due to black hole production in several ways. First,
although a large rate may be attributed to either an enhanced flux or
an enhanced black hole cross section, these possibilities may be
distinguished by searches for Earth-skimming
neutrinos~\cite{Feng:2001ue,Bertou:2001vm}.  While an enhanced flux
increases these rates, a large black hole cross section will suppress
them, since the hadronic decay products of black hole evaporation will
not escape the Earth's crust.

Second, showers from black hole production have distinctive
characteristics.  In the SM, typical hadronic showers, as initiated by
nucleons or nuclei, occur high in the atmosphere.  Deep atmospheric
showers arise only from $\nu N \to \ell X$, resulting in a hadronic
shower initiated by the struck quark, possibly accompanied by an
electromagnetic shower carrying most of the incident energy, depending
on the neutrino flavor. Black hole events are markedly different.  The
black hole rest lifetime is $\tau \sim (1/\mstar)
(\mbh/\mstar)^{(3+n)/(1+n)}$.  Since $\mstar^{-1} \sim \tev^{-1} \sim
10^{-27}~\s$, even the largest black holes produced evaporate
effectively instantaneously.  In contrast to SM showers, however,
black hole showers have small electromagnetic components, and the
average multiplicity in black hole decays
is~\cite{Giddings:2001bu,Dimopoulos:2001hw}
\begin{equation}
\langle N \rangle \approx \frac{\mbh}{2 T_H}
= \frac{2 \sqrt{\pi}}{1+n}
\Biggl[ \frac{\mbh}{\mstar} \Biggr]^{\frac{2+n}{1+n}}
\left[ \frac{8 \Gamma \left(\frac{3+n}{2} \right)}
{2+n} \right]^{\frac{1}{1+n}} ,
\label{averageN}
\end{equation}
where $T_H$ is the Hawking temperature.  Large mass black holes
therefore decay to large numbers of quarks and gluons, and black hole
showers will appear more nucleus-like than SM events, with the
discrepancy growing with black hole mass.  Nucleus showers differ from
nucleon showers in several ways~\cite{Nagano:2000ve}.  $\xmax$, the
atmospheric depth at which the number of particles in a shower reaches
its maximum, is significantly lower for nuclei, and shower-to-shower
fluctuations in $\xmax$ and the number of electrons are also
smaller. Black holes and SM events may therefore be distinguished
based on shower characteristics, at least on a statistical basis. Note
from Fig.~\ref{fig:nmbhmin} that a fairly smooth distribution of black
hole masses is expected.  If large numbers of black holes are found,
the correlations of shower energy with $\mbh$ and $\xmax$ with
$\langle N \rangle$ will also allow tests of Hawking evaporation and
possibly even measurements of $n$ and $\mstar$ through
\eqref{averageN}.

Before closing, we comment on the possible relevance of black hole
production to the GZK paradox.  As noted above, the cross sections of
Fig.~\ref{fig:sig1} may be enhanced, especially for $\mbh \sim
\mstar$, where the behavior of black holes and related objects is very
poorly understood.  In addition, if effectively four-dimensional black
holes are produced, as may be possible in warped scenarios with small
curvature scales, we find cross sections of 10 mb for $E_{\nu} \sim
10^{12}~\gev$.  If these or other enhancements are large enough to
bring the cross sections to the 100 mb level, cosmic neutrinos will be
primaries immune to GZK-type cutoffs that produce hadronic showers
high in the atmosphere, providing a viable resolution to the GZK
puzzle. While the required enhancement is large and speculative, the
qualitative merits of black hole production as a solution to the GZK
paradox are suggestive and deserve further study.

To summarize, in TeV-scale gravity scenarios, ultra-high energy cosmic
neutrinos will produce black holes in the Earth's atmosphere, leading
to anomalously large rates for quasi-horizontal hadronic showers.  If
the LHC is to be a black hole factory, at least tens to hundreds of
black holes will be detected at the Auger Observatory before the LHC
begins operation.  Such events are powerful probes of extra
dimensions, and may provide information about black holes in the late
stages of evaporation.

\begin{acknowledgments}
We thank S.~Giddings, T.~Han, G.~Horowitz, F.~Wilczek, and especially
P.~Argyres for discussions.  ADS thanks the Center for Theoretical
Physics at MIT for hospitality and support. This work was supported in
part by the Department of Energy under cooperative research agreement
DF-FC02-94ER40818.  The work of ADS is supported in part by DOE Grant
No.\ DE-FG01-00ER45832 and NSF Grant No.\ PHY-0071312.
\end{acknowledgments}

%%%%%%%%%%%%%%%%%%%%%%%%%%%%%%%%%%%%%%%%%%%%%%%%%%%%%%

%%%%%%%%%%%%%%%%%%%%%%%%%%%%%%%%%%%%%%%


\begin{thebibliography}{99}
%%%%%%%%%%%%%%%%%%%%%%%%%%%%%%%%%%%%%%%%%%%%%%%%%%%%%%

\bibitem{Hawking:1975sw}
S.~W.~Hawking,
%``Particle Creation By Black Holes,''
Commun.\ Math.\ Phys.\  {\bf 43}, 199 (1975).
%%CITATION = CMPHA,43,199;%%

\bibitem{Banks:1999gd}
T.~Banks and W.~Fischler,
%``A model for high energy scattering in quantum gravity,''
hep-th/9906038.
%%CITATION = HEP-TH 9906038;%%

\bibitem{Emparan:2000rs}
R.~Emparan, G.~T.~Horowitz and R.~C.~Myers,
%``Black holes radiate mainly on the brane,''
Phys.\ Rev.\ Lett.\  {\bf 85}, 499 (2000)
[hep-th/0003118].
%%CITATION = HEP-TH 0003118;%%.  

\bibitem{Giddings:2001bu}
S.~B.~Giddings and S.~Thomas,
%``High energy colliders as black hole factories: The end of short 
%distance physics,''
hep-ph/0106219.
%%CITATION = HEP-PH 0106219;%%

\bibitem{Dimopoulos:2001hw}
S.~Dimopoulos and G.~Landsberg,
%``Black holes at the LHC,''
hep-ph/0106295.
%%CITATION = HEP-PH 0106295;%%

\bibitem{Feng:2001ue}
J.~L.~Feng, P.~Fisher, F.~Wilczek and T.~M.~Yu,
%``Observability of earth-skimming ultra-high energy neutrinos,''
hep-ph/0105067;
%%CITATION = HEP-PH 0105067;%%
%\bibitem{Kusenko:2001gj}
A.~Kusenko and T.~Weiler,
%``Neutrino cross sections at high energies and the future
%observations of ultrahigh-energy cosmic rays,''
hep-ph/0106071.
%%CITATION = HEP-PH 0106071;%%

\bibitem{Bertou:2001vm}
X.~Bertou, P.~Billoir, O.~Deligny, C.~Lachaud and A.~Letessier-Selvon,
%``Tau neutrinos in the Auger observatory: A new window to UHECR
%sources,''
astro-ph/0104452.
%%CITATION = ASTRO-PH 0104452;%%

\bibitem{Capelle:1998zz}
K.~S.~Capelle, J.~W.~Cronin, G.~Parente and E.~Zas,
%``On the detection of ultra high energy neutrinos with the Auger
%Observatory,''
Astropart.\ Phys.\ {\bf 8}, 321 (1998)
[astro-ph/9801313].
%%CITATION = ASTRO-PH 9801313;%%

\bibitem{Coutu:1999ub}
S.~Coutu, X.~Bertou and P.~Billoir [AUGER Collaboration],
%``Ultrahigh energy neutrinos with AUGER,''
%in {\em Proceedings of the 23rd Johns Hopkins Workshop on Current
%Problems in Particle Theory}, Baltimore (1999), 
Auger Note GAP-1999-030.

%J.~C.~Diaz, M.~G.~do Amaral, and R.~C.~Shellard, 
%%``Weakly interacting particles viewed by fluorescence detectors,'' 
%Auger Note GAP-2000-058;
\bibitem{Diaz:2001mv}
J.~C.~Diaz, R.~C.~Shellard and M.~G.~Amaral,
%``The acceptance of fluorescence detectors for quasi horizontal 
%showers induced by weak interacting particles,''
Nucl.\ Phys.\ Proc.\ Suppl.\  {\bf 97}, 247 (2001).
%%CITATION = NUPHZ,97,247;%%

\bibitem{Antoniadis:1998ig}
I.~Antoniadis, N.~Arkani-Hamed, S.~Dimopoulos and G.~Dvali,
%``New dimensions at a millimeter to a Fermi and superstrings at a TeV,''
Phys.\ Lett.\ B {\bf 436}, 257 (1998)
[hep-ph/9804398].
%%CITATION = HEP-PH 9804398;%%

\bibitem{Peskin:2000ti}
M.~E.~Peskin,
%``Theoretical summary,''
hep-ph/0002041.
%%CITATION = HEP-PH 0002041;%%

\bibitem{Randall:1999ee}
L.~J.~Randall and R.~Sundrum,
%``A large mass hierarchy from a small extra dimension,''
Phys.\ Rev.\ Lett.\  {\bf 83}, 3370 (1999)
[hep-ph/9905221].
%%CITATION = HEP-PH 9905221;%%
 
\bibitem{Davoudiasl:2000jd}
H.~Davoudiasl, J.~L.~Hewett and T.~G.~Rizzo,
%``Phenomenology of the Randall-Sundrum gauge hierarchy model,''
Phys.\ Rev.\ Lett.\  {\bf 84}, 2080 (2000)
[hep-ph/9909255].
%%CITATION = HEP-PH 9909255;%%

\bibitem{Nussinov:1999jt}
See, \eg, 
S.~Nussinov and R.~Shrock,
%``Some remarks on theories with large compact dimensions and
%TeV-scale  quantum gravity,''
Phys.\ Rev.\ D {\bf 59}, 105002 (1999)
[hep-ph/9811323];
%%CITATION = HEP-PH 9811323;%%
%\bibitem{Jain:2000pu}
P.~Jain, D.~W.~McKay, S.~Panda and J.~P.~Ralston,
%``Extra dimensions and strong neutrino nucleon interactions above
%10**19-eV: Breaking the GZK barrier,''
Phys.\ Lett.\ B {\bf 484}, 267 (2000)
[hep-ph/0001031];
%%CITATION = HEP-PH 0001031;%%
%\bibitem{Tyler:2001gt}
C.~Tyler, A.~V.~Olinto and G.~Sigl,
%``Cosmic neutrinos and new physics beyond the electroweak scale,''
Phys.\ Rev.\ D {\bf 63}, 055001 (2001)
[hep-ph/0002257];
%%CITATION = HEP-PH 0002257;%%
%\bibitem{Alvarez-Muniz:2001mk}
J.~Alvarez-Muniz, F.~Halzen, T.~Han and D.~Hooper,
%``Phenomenology of high energy neutrinos in low-scale quantum gravity 
%models,''
hep-ph/0107057.
%%CITATION = HEP-PH 0107057;%%

\bibitem{Myers:1986un}
R.~C.~Myers and M.~J.~Perry,
%``Black Holes In Higher Dimensional Space-Times,''
Annals Phys.\  {\bf 172}, 304 (1986);
%%CITATION = APNYA,172,304;%%
%\bibitem{Argyres:1998qn}
P.~C.~Argyres, S.~Dimopoulos and J.~March-Russell,
%``Black holes and sub-millimeter dimensions,''
Phys.\ Lett.\ B {\bf 441}, 96 (1998)
[hep-th/9808138].
%%CITATION = HEP-TH 9808138;%%

\bibitem{D'Eath:1993gr}
P.~D.~D'Eath,
%``Gravitational radiation in high speed black hole collisions,''
Class.\ Quant.\ Grav.\  {\bf 10}, S207 (1993).
%%CITATION = CQGRD,10,S207;%%

\bibitem{Lai:2000wy}
H.~L.~Lai {\it et al.}  [CTEQ Collaboration],
%``Global {QCD} analysis of parton structure of the nucleon: CTEQ5 
%parton distributions,''
Eur.\ Phys.\ J.\ C {\bf 12}, 375 (2000)
[hep-ph/9903282].
%%CITATION = HEP-PH 9903282;%%

\bibitem{Greisen:1966jv}
K.~Greisen,
%``End To The Cosmic Ray Spectrum?,''
Phys.\ Rev.\ Lett.\ {\bf 16}, 748 (1966).
%%CITATION = PRLTA,16,748;%%

\bibitem{Stecker:1979ah}
F.~W.~Stecker,
%``Diffuse Fluxes Of Cosmic High-Energy Neutrinos,''
Astrophys.\ J.\  {\bf 228}, 919 (1979).
%%CITATION = ASJOA,228,919;%%

\bibitem{Hill:1985mk}
C.~T.~Hill and D.~N.~Schramm,
%``The Ultrahigh-Energy Cosmic Ray Spectrum,''
Phys.\ Rev.\ D {\bf 31}, 564 (1985);
%%CITATION = PHRVA,D31,564;%%
%\bibitem{Protheroe:1996ft}
R.~J.~Protheroe and P.~A.~Johnson,
%``Propagation of ultrahigh-energy protons over cosmological distances
%and implications for topological defect models,''
Astropart.\ Phys.\  {\bf 4}, 253 (1996)
[astro-ph/9506119].
%%CITATION = ASTRO-PH 9506119;%%

\bibitem{Yoshida:1997ie}
S.~Yoshida, H.~Dai, C.~C.~Jui and P.~Sommers,
%``Extremely high energy neutrinos and their detection,''
Astrophys.\ J.\ {\bf 479}, 547 (1997)
[astro-ph/9608186].
%%CITATION = ASTRO-PH 9608186;%%

\bibitem{Zatsepin:1966jv}
G.~T.~Zatsepin and V.~A.~Kuz'min,
%``Upper Limit Of The Spectrum Of Cosmic Rays,''
JETP Lett.\ {\bf 4}, 78 (1966).
%%CITATION = JTPLA,4,78;%%

\bibitem{Voloshin:2001vs}
M.~B.~Voloshin,
%``Semiclassical suppression of black hole production in particle  
% collisions,''
hep-ph/0107119.
%%CITATION = HEP-PH 0107119;%%

\bibitem{Dimopoulos:2001qe}
See, \eg,
S.~Dimopoulos and R.~Emparan,
%``String balls at the LHC and beyond,''
hep-ph/0108060.
%%CITATION = HEP-PH 0108060;%%

\bibitem{Stecker:2000ek}
See, \eg,
F.~W.~Stecker,
%``The curious adventure of the ultrahigh energy cosmic rays,''
astro-ph/0101072.
%%CITATION = ASTRO-PH 0101072;%%.

\bibitem{Nagano:2000ve}
For a review, see
M.~Nagano and A.~A.~Watson,
%``Observations and implications of the ultrahigh-energy cosmic rays,''
Rev.\ Mod.\ Phys.\  {\bf 72}, 689 (2000).
%%CITATION = RMPHA,72,689;%%.  

%%%%%%%%%%%%%%%%%%%%%%%%%%%%%%%%%%%%%%%
\end{thebibliography}
\end{document}